\documentclass[journal]{IEEEtran}
\usepackage{url}
\usepackage{graphicx}
\usepackage{cite}
\usepackage{paralist}
\usepackage{booktabs}
\usepackage{threeparttable}
\usepackage{multirow}
\usepackage{stfloats}
\usepackage{float}
\usepackage{wasysym}
\usepackage{graphicx}
\usepackage{subcaption}
\usepackage{xcolor}
\usepackage[hidelinks]{hyperref}

\begin{document}

\title{Adaptive Time Budgets for Safe and Comfortable Vehicle Control Transition in Conditionally Automated Driving}

\author{Kexin~Liang,
        Simeon C.~Calvert,
        and~J.W.C.~van Lint
\thanks{K. Liang, S.C. Calvert, and J.W.C. van Lint are with the Department
of Transport and Planning, Faculty of Civil Engineering and Geosciences, Delft University of Technology.}}

%

\maketitle

\begin{abstract}

Conditionally automated driving requires drivers to resume vehicle control promptly when automation reaches its operational limits. Ensuring smooth vehicle control transitions is critical for the safety and efficiency of mixed-traffic transportation systems, where complex interactions and variable traffic behaviors pose additional challenges. This study addresses this challenge by introducing an adaptive time budget framework that provides drivers with sufficient time to complete takeovers both safely and comfortably across diverse scenarios. We focus in particular on the takeover buffer—the extra time available after drivers consciously resume control to complete evasive maneuvers. A driving simulator experiment is conducted to evaluate the influence of different takeover buffer lengths on safety-related indicators (minimum time-to-collision, maximum deceleration, and steering wheel angle) and subjective assessments (perceived time sufficiency, perceived risk, and performance satisfaction). Results show that (i) takeover buffers of about 5–6 seconds consistently lead to optimal safety and comfort; and (ii) drivers prefer relatively stable takeover buffers across varying traffic densities and $n$-back tasks. This study introduces an adaptive time budget framework that dynamically allocates transition time by incorporating a predicted takeover time and a preferred takeover buffer (piece-wise function). This can serve as an important first step toward providing drivers with sufficient time to resume vehicle control across diverse scenarios, which needs to be validated in more diverse and real-world driving contexts. By aligning the provided time budget with driver needs under specific circumstances, the adaptive framework can improve reliability of control transitions, facilitate human-centered automated driving, reduce crash risk, and maintain overall traffic efficiency.




\end{abstract}


\IEEEpeerreviewmaketitle

\section{Introduction}

\IEEEPARstart{C}{onditionally} automated driving systems issue takeover requests (TORs) in situations that exceed their operational capabilities, requiring drivers to promptly resume manual control and maintain safe vehicle operation. A key factor in ensuring the smoothness of such control transitions is the time budget, i.e., the time offered by automation for control transitions. When the time budget is too short to accommodate the required takeover time (ToT, the time drivers need to regain manual vehicle control after receiving a TOR \cite{ISO21959}), the risk of accidents increases \cite{gold2013take} as drivers may lack adequate time to perceive, assess, and respond to the situation. Conversely, time budgets that substantially exceed the required ToT may also introduce risks: such TORs can be perceived as false alarms \cite{huang2022takeover, skrickij2020autonomous}, leading to reduced driver attention and potential dangers. Therefore, defining and allocating sufficient time budgets is essential to ensure driving safety and user experience in vehicle control transitions.

Previous research on time budget determination has primarily focused on identifying fixed time budgets for safe driver takeover. A widely cited time budget for drivers to resume vehicle control is 7s \cite{gold2013take}, which has been adopted in numerous takeover experiments \cite{gold2016taking, korber2016influence, jarosch2019effects, li2021drivers}. In on-road applications, Mercedes-Benz, for example, provides drivers with 10s to resume control \footnote{\url{https://abcnews.go.com/Business/future-driving-hands-off-eyes-off-mercedes-benz/story?id=103681808}}. However, this fixed-time-budget approach overlooks the substantial variability in the required ToT across drivers and scenarios \cite{zhang2021optimal, zhang2023determine}. Favarò, Eurich, and Rizvi \cite{favaro2019human} reported 1,143 takeover cases from four vehicle manufacturers and found ToT values ranging from 0.83 s to 3.10 s. Zhang et al. \cite{zhang2019determinants} reported ToT values ranging from 0.69 s to 19.79 s across 520 takeovers, with a long right-tailed distribution. These findings highlight the limitations of fixed time budgets in accommodating such variability of ToT, as they may be insufficient for drivers or situations requiring longer ToT or unnecessarily long for others, potentially leading to safety risks and reduced system efficiency in complex traffic scenarios. Determining how to provide sufficient time budgets for different drivers to take over vehicle control smoothly across diverse scenarios remains a key challenge.

Providing adaptive time budgets is a promising approach for addressing the limitations of fixed time allocations by dynamically adjusting the provided time to suffice the required ToT across diverse drivers and driving scenarios \cite{li2021adaptive, liang2025predicting}. One of the practical foundations for determining adaptive time budgets comes from Marberger et al. (2017), which proposed using the difference between the available time budget and the time required for a safe takeover to quantify driver availability \cite{marberger2017understanding}. In our previous review study \cite{liang2025towards}, a similar concept of ``takeover buffer'' was defined as the surplus time remaining within the provided time budget after subtracting the driver’s required ToT. Thus, a sufficient time budget can be determined by adding the predicted ToT required by drivers to the optimal takeover buffer. Among these two components, prior work has predominantly concentrated on predicting ToT \cite{liu2025takeover, ayoub2022predicting}. However, limited effort has been devoted to determining takeover buffers. Designing appropriately sized takeover buffers tailored to situational demands has not yet been thoroughly investigated \cite{tanshi2022determination, gold2018modeling}.



This study aims to explore the optimal allocation of takeover buffer by investigating its relationship with takeover time and performance. Specifically, we focus on two main questions:
\begin{itemize}
    \item How does the takeover buffer affect drivers’ response quality and subjective experience across diverse takeover situations?
    \item What durations of takeover buffer do drivers prefer when their required takeover time varies?
\end{itemize}

To answer these questions, we conducted a driving simulator experiment where drivers performed takeover maneuvers across nine scenarios (three traffic densities $*$ three non-driving related tasks). With the data collected from the experiment, we conduct a comprehensive analysis of the effects of takeover buffer on various aspects of takeover performance, including (i) subjective driver experience, encompassing performance satisfaction, perceived time sufficiency,  and perceived risk; and (ii) objective response quality, encompassing minimum time to collision, maximum deceleration, and maximum steering wheel angle during takeovers. We also investigate drivers' preferred time budget and takeover buffer across various takeover times. Based on the findings, we propose an adaptive time budget strategy that sets the time budget as the predicted takeover time plus a saturated takeover buffer. The findings provide actionable guidance for designing sufficient time budgets that support safe and comfortable takeovers, thereby informing the development of more effective human–vehicle interaction strategies for control transitions in conditionally automated driving.



\section{Related work}
\label{sec: related work}

Research on the takeover buffer and its relationship with takeover time (ToT) and performance is currently underexplored. Li et al. \cite{li2021drivers} introduced a concept similar to the takeover buffer, termed ``time to boundary at takeover timing'' (TTBT), and analyzed its relationship with crash occurrence using logistic regression. Their results showed that longer TTBT values were linked to a lower probability of crashes. In their following research, Li et al. \cite{li2021adaptive} offered participants four TTBT options (3s, 4s, 5s, and 6s). While all takeovers were successfully completed (meeting the minimum safety requirement), the 4-second option emerged as the most preferred. This aligns with \cite{zhang2025effects}, where driver satisfaction showed an inverted U-shape with interval length. A 7s interval was preferred for sufficient situation awareness, good performance, and high satisfaction, while 9s led to more attention toward the non-driving-related task. These results indicate a potentially consistent optimal takeover buffer that balances safety and driver preference; however, its stability across different takeover times and the factors underlying this consistency remain to be investigated.

From a finer-grained perspective, the takeover buffer consists of two elements: safety buffer and comfort buffer. The \textit{Safety Buffer} represents the temporal margin from when the driver regains conscious control of the vehicle to the point where evasive actions are no longer required to avoid a potential collision. This concept aligns with \textit{Time to Control (TC)}, defined as the time from receiving a takeover request to the moment when no further deceleration is necessary to avoid a hazard \cite{papadimitriou2024method}. The safety buffer can thus be calculated as the difference between TC and the ToT, and has been recognized as a key factor in determining adaptive time budgets for successful takeovers \cite{marberger2017understanding}. Specifically, the allocated time budget should exceed the sum of the predicted ToT and the necessary safety buffer \cite{tanshi2022determination}, thus providing a sufficient margin for safe maneuver execution. Beyond this minimum safety requirement, adding an extra \textit{Comfort Buffer}, i.e., a psychological margin during which no immediate action is required, can enhance user experience by allowing drivers to feel less stressed. This formulation parallels the two-stage takeover request framework, where the first warning serves as a preparatory signal that does not require immediate action (analogous to the comfort buffer), while the second warning prompts drivers to execute actual maneuvers (corresponding to the sum of ToT and safety buffer) \cite{zhang2021optimal}. In the study of Papadimitriou et al. \cite{papadimitriou2024method}, a comfort buffer of 0.9s was used as the minimum threshold, whereas values below 0.9s were considered insufficient. Note that a longer comfort buffer does not necessarily improve driver experience: both Li et al. \cite{li2021adaptive} and Zhang et al. \cite{zhang2025effects} found that driver satisfaction exhibits an inverted U-shaped relationship with buffer times. How to allocate optimal takeover buffer while balancing safety and comfort buffers remains underexplored, highlighting the need for empirical studies to guide adaptive time budget design.

\begin{figure}[!h]
    \centering
    \includegraphics[width=\columnwidth]{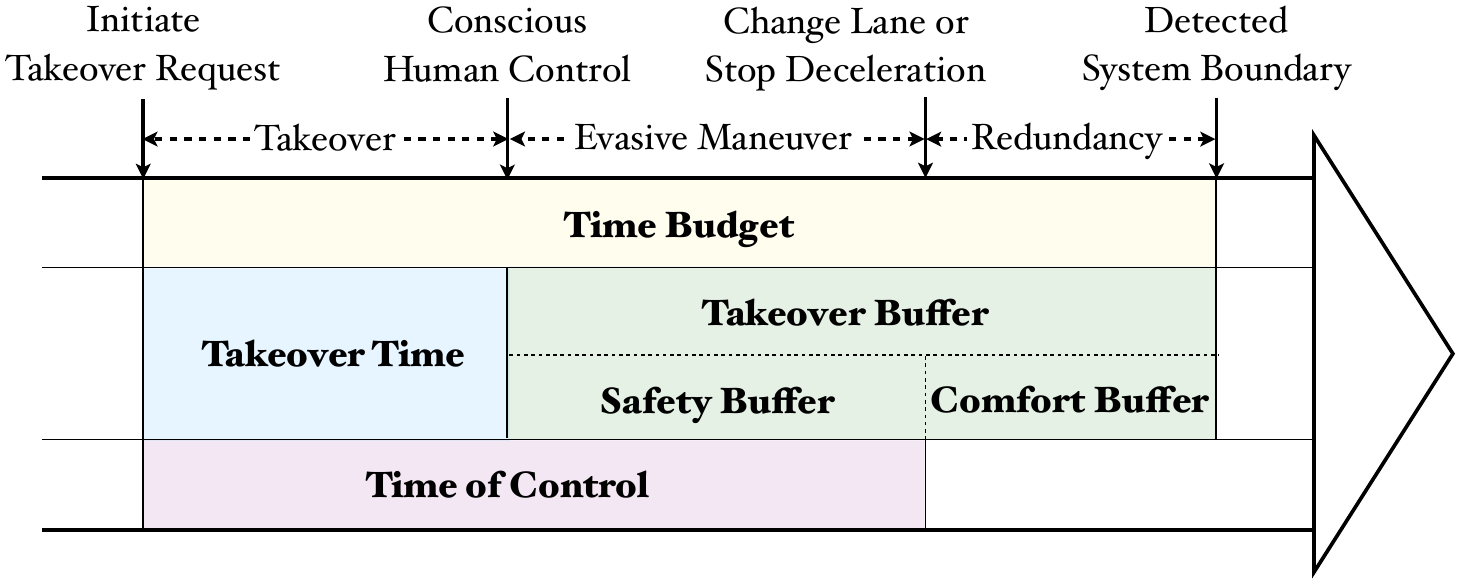}
    \caption{Temporal metrics of drivers' takeovers: a timeline-based overview (Segment lengths and boundaries are schematic and may vary with context).}
    \label{fig: time metrics}
\end{figure}

To distinguish key temporal metrics associated with the driver takeover process, Figure \ref{fig: time metrics} presents these metrics along a typical takeover timeline, clarifying their definitions, interrelationships, and sequential order. This study focuses on the takeover buffer and its subcomponents, examining how they vary with takeover times and influence various takeover performance metrics. The findings can be used to inform the design of adaptive time budgets that meet diverse driver and scenario demands.

\section{Method}
\label{sec: method}

\subsection{Driving simulator experiment}
\label{subsec: experiment}

\subsubsection{Experiment setup}

\begin{figure}[ht]
    \centering
    \includegraphics[width=\columnwidth]{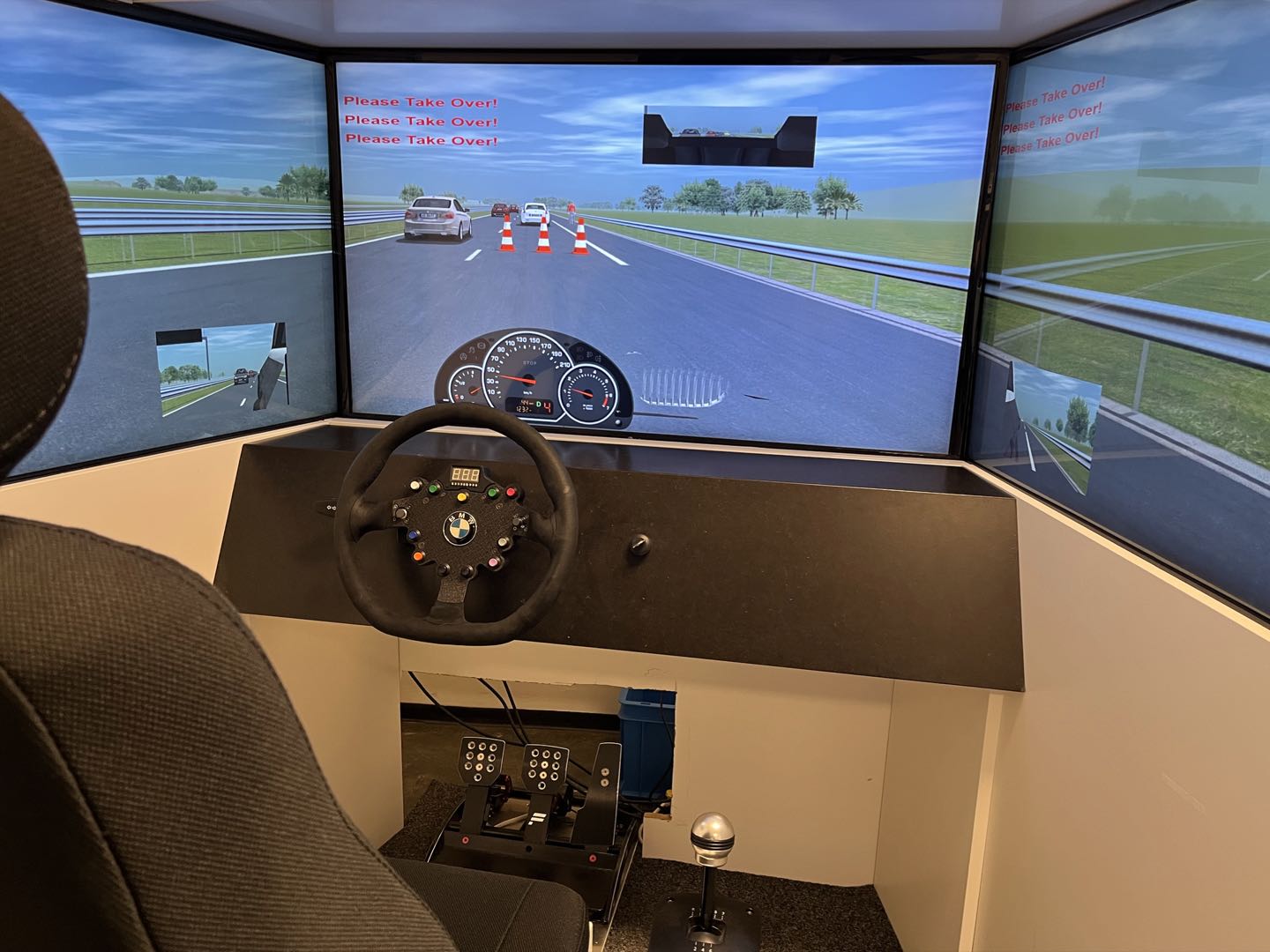} 
    \caption{Driving simulator at TU Delft.}
    \label{subfig: simulator} 
\end{figure}

The experiment took place at Delft University of Technology (TU Delft) using a fixed-base, medium-fidelity driving simulator. The simulator, depicted in Figure~\ref{subfig: simulator}, features three 4k-resolution screens providing views from the windshield and two side windows. Experiment scenarios are programmed on a Windows 10 PC.

This study simulates a two-lane motorway driving environment. The automated driving speed is set to 100 km/h, adhering to the daytime speed limit on motorways in the Netherlands. A takeover request is triggered when the time headway between the ego vehicle and the collision ahead reaches 7s \cite{deniel2024driver, manchon2023calibration}. This request consists of three auditory beeps accompanied by the display of the message ``Please Take Over!'' shown in three lines of text at the top left corner of the windshield (see Figure~\ref{subfig: simulator}).

To investigate drivers' responses to takeover requests in different scenarios, we employ a three (traffic densities) $\times$ three (non-driving-related tasks) repeated measures design. Traffic densities are set at $0$, $10$, and $20$ vehicles per kilometer, while $n$-back tasks with $n = 0, 1, 2$ serve as the non-driving-related tasks to impose differing levels of mental demand. This $n$-back task involves participants tracking the position of a blue box and pressing a button when the current position is the same as the one occurring $n$ steps back in the sequence \cite{liang2024examining, liang2025predicting}. The nine takeover scenarios are arranged using a Latin Square design \cite{calvert2014application} to minimize order and learning effects. After receiving a takeover request, participants are required to immediately detach from the $n$-back task and begin resuming vehicle control from automation.

\subsubsection{Experiment procedure}

Before the experiment, participants were briefed on the abilities and boundaries of the conditionally automated driving system (CADS), as well as on the $n$-back task. A ten-minute practice drive was provided for participants to get familiar with the simulator and the takeover process, which helps to reduce learning effects. 

\begin{figure*}[!h]
    \centering
    \includegraphics[width=\textwidth]{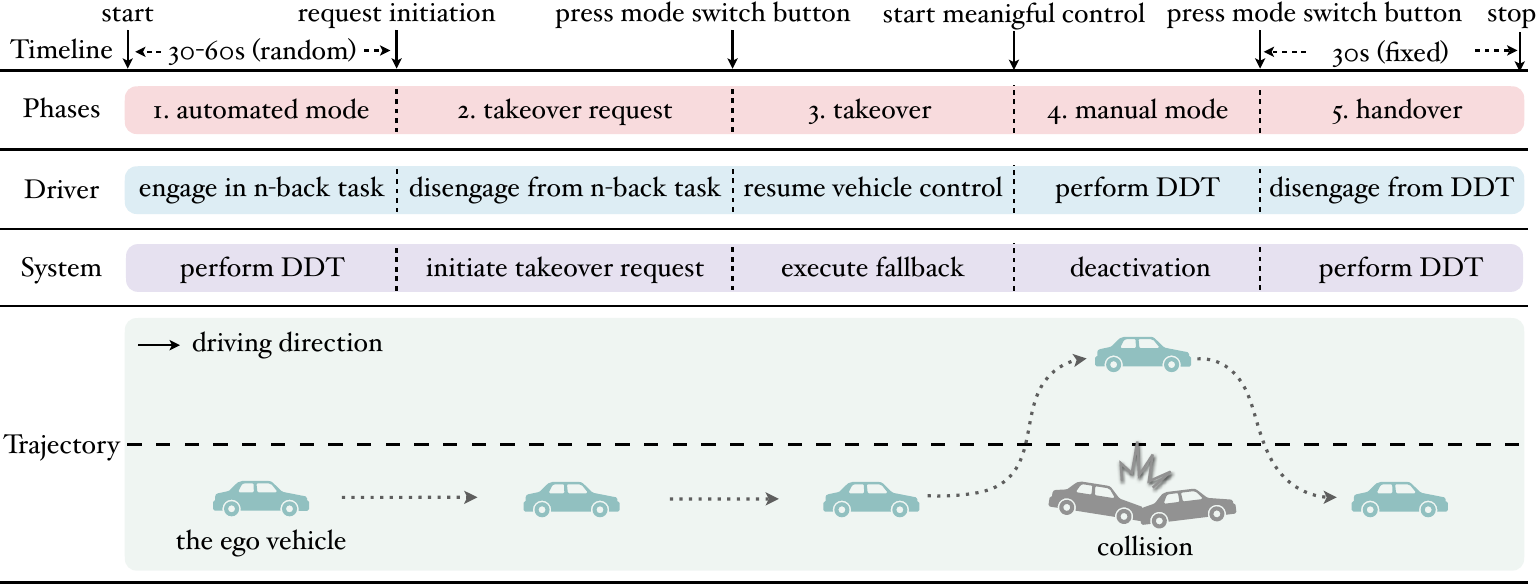}
    \caption{The procedure of a takeover event (DDT: Dynamic Driving Task) \cite{liang2025predicting}.}
    \label{fig: takeover session}
\end{figure*}

During the experiment, participants experienced nine takeover events. Each takeover event includes five phases as shown in Figure~\ref{fig: takeover session} : \begin{inparaenum}[(i)]
        \textit{\item Automated mode:} The takeover event starts from automated mode while participants are engaged in the $n$-back task.
        \textit{\item Takeover request:} The takeover request is randomly initiated between 30s and 60s after entering the automated mode. This time window ensures participants have sufficient time to engage with the $n$-back task before the request, as well as to take over and stabilize the ego vehicle after the request. Randomizing the timing of takeover requests aims to eliminate the predictability associated with fixed durations in automated mode.
        \textit{\item Takeover:} Participants are instructed to promptly detach from the $n$-back task and begin resuming control of the ego vehicle upon receiving the request.
        \textit{\item Manual mode:} After regaining conscious control of the ego vehicle, participants are tasked with an evasive maneuver, which involves pulling out to the left lane, overtaking the detected collision ahead, and pulling over to the right-hand lane after bypassing the collision.
        \textit{\item Handover:} Participants are instructed to hand over vehicle control back to the CADS once they believe it is safe to do so after stabilizing the vehicle on the right-hand lane.
    \end{inparaenum}

After each takeover event, participants take a break and report their subjective feelings of the last takeover, including \begin{inparaenum}[(i)]
\item \textit{performance satisfaction}: participants rated their agreement with “I was satisfied with my performance in taking over car control” using a five-point Likert scale (1 = Strongly Disagree, 5 = Strongly Agree).
\item \textit{perceived time sufficiency}: participants responded to: “To complete the required bypass maneuvers safely and comfortably, how much more or less time would you prefer for the takeover?” on a sliding scale from -10 to +10 seconds. Values were normalized to [0, 1], where 0 = Inadequate, 1 = Excessive.
\item \textit{perceived risk}: participants rated their agreement with “I was worried about being involved in a traffic accident during the takeover process” on a five-point Likert scale (1 = Strongly Disagree, 5 = Strongly Agree). 
\end{inparaenum}

\subsection{Data acquisition and analysis}
\label{subsec: data analysis}

\subsubsection{Participant sample}

In total, 57 participants took part in the study. Gender and age distributions are fairly balanced, with 33 males and 24 females, and participants spanning a broad age range (mean = 38.51, SD = 17.23). Their individual traits were assessed using the driver characteristic questionnaire detailed in our previous work \cite{liang2025predicting}, and summarized through descriptive statistics in Table~\ref{tab: driver characteristics}. The sample comprises a heterogeneous group in terms of demographic profiles, skill-related attributes, and style-related traits. This diversity enables a comprehensive analysis of the associations between takeover-related intervals and performance outcomes, thereby providing valuable insights for the development of adaptive time budget strategies.

\begin{table*}[!h]
\centering
\caption{Descriptive statistics of participants' driver characteristics.}
\label{tab: driver characteristics}
\resizebox{\textwidth}{!}{ 

\begin{tabular}{p{14em} p{3em}p{3em} p{2em} p{3em}  p{10em} p{2em}p{2em} p{2em} p{2em}}

\toprule
\textbf{characteristics} &  \textbf{mean} &\textbf{SD} &  \textbf{min} &\textbf{max} & \textbf{characteristics} & \textbf{mean} & \textbf{SD}&  \textbf{min} &\textbf{max} \\ \midrule
age & 37.47 & 16.58& 18 & 72& takeover skill  &3.73  & 0.55 & 2.56& 4.94\\
gender & 0.42 & 0.50 & 0 & 1& risk-taking attitude  &2.18  & 0.49& 1& 3.2\\
accumulated driving years & 17.04 & 16.91 & 1 & 55& trust in automation  & 3.98& 0.88& 2& 5\\
accumulated driving kilometers& 5840.91 & 6936.30& 0& 30000&  takeover style - reckless  &2.57 & 0.73& 1&4.4 \\
driving frequency  &0.30 & 0.27 & 0& 1& takeover style - anxious & 2.55& 0.69& 1& 4.2	\\
driving skill  & 1.40 & 0.66& 0& 2& takeover style - angry  &  1.96& 0.60& 1& 3.2\\
driver assistance usage frequency & 0.30&0.37& 0& 1&  takeover style - patient & 3.42& 0.49& 2.4 &4.4 \\
                                 \bottomrule
\end{tabular}	}							

\begin{tablenotes}
    \footnotesize
    \item $*$ For gender: Male = 0; Female = 1.
    \item $*$ For driving skill: Inexperienced = 0; Intermediate = 1; Experienced = 2.
    \item $*$ Characteristics on the right side of the table (including takeover skill, risk-taking attitude, trust in automation, takeover style - reckless, takeover style - anxious, takeover style - angry, and takeover style - patient) are assessed on five-point scales ranging from 1 to 5.
\end{tablenotes}
\end{table*}

\subsubsection{Collected variables} 

To explore the optimal takeover buffer for drivers resuming vehicle control from CADS, this study records data from the initiation of takeover requests until drivers successfully change lanes to avoid collisions in distinct scenarios. The collected data can be classified into temporal metrics and performance metrics, as detailed in Table \ref{tab: data overview}.

\begin{table*}[!h]
\centering
\caption{Overview of collected data.}
\label{tab: data overview}
\resizebox{\textwidth}{!}{
\begin{tabular}{p{5em} p{12em} l p{32em}}
\toprule
\textbf{Category} & \textbf{Metric} & \textbf{Unit} & \textbf{Description} \\ 
\midrule

\multirow{17}{5em}{Temporal intervals} 
& Time Budget & $s$ & Time to collision (TTC) when a Takeover Request (TOR) is issued. \\
& Takeover Time & $s$ & Time from the TOR to the driver's first conscious operational response. \\
& Takeover Buffer & $s$ & Remaining time within the time budget after the driver resumes conscious control; calculated as Time Budget - Takeover Time. \\
& Time to Control & $s$ & Time to complete evasive maneuver, defined as the minimum of the time to lane change and time to stop deceleration \cite{papadimitriou2024method}. \\
& Safety Buffer & $s$ & Time margin for completing necessary intervention (defined as completing deceleration, initiating acceleration, or finishing a lane change in this study), calculated as Time to Control - Takeover Time. \\
& Comfort Buffer & $s$ & Remaining time within time budget after completing evasive actions for optimizing driver experience; calculated as Time Budget - Time to Control. \\
& Preferred Time Adjustment & $s$ & Participants responded to: “To complete the required bypass maneuvers safely and comfortably, how much more or less time would you prefer for the takeover?” on a sliding scale from -10 to +10 seconds.\\
& Preferred Time Budget & $s$ & = Time Budget + Preferred Time Adjustment. \\
& Preferred Takeover Buffer & $s$ & = Preferred Time Budget - Takeover Time \\
\midrule

\multirow{11}{5em}{Performance metrics} 
& Performance Satisfaction & -- & Participants rated their agreement with “I was satisfied with my performance in taking over car control” using a five-point Likert scale (1 = Strongly Disagree, 5 = Strongly Agree). \\
& Time Sufficiency & -- & Normalize the values of Preferred Time Adjustment to a range of [0, 1], where 0 = Inadequate, 1 = Excessive. \\
& Perceived Risk & -- & Participants rated their agreement with “I was worried about being involved in a traffic accident during the takeover process” on a five-point Likert scale (1 = Strongly Disagree, 5 = Strongly Agree). \\
& Min TTC & $s$ & The minimum time to collision from the TOR to the lane change. \\
& Max Deceleration & $m/s^2$ & Maximum deceleration observed between the TOR and the lane change. \\
& Max Steering Wheel Angle & $rad$ & Maximum steering angle recorded between the TOR and the lane change. \\
\bottomrule
\end{tabular}}
\end{table*}

\subsubsection{Data analysis}

The statistical analyses are conducted in Python (version 3.11), consisting of four main stages: First, the distributions of (objective and subjective) performance metrics across traffic densities and non-driving-related tasks are visualized using boxplots in Section \ref{subsec: takeover performance}. This helps to assess the effectiveness of fixed time budgets (7s in this study) across various takeover scenarios. Then, the relationships between takeover buffer levels—including its components, safety buffer and comfort buffer—and various takeover performance metrics are examined in \ref{subsec: takeover buffer}. This analysis aims to determine how buffer allocation influences takeover safety and user comfort. Next, preferred time budgets and preferred takeover buffers are compared across different takeover time levels in Section \ref{subsec: driver preference} to identify the factors driving variations in drivers’ temporal preferences. Finally, the findings from the above analyses are integrated into the design of an adaptive time budget framework in Section \ref{subsec: adaptive time budget}. Its validity is assessed by comparing the proposed adaptive time budget with the fixed 7s time budget, evaluating which more accurately reflects drivers’ preferred temporal allocations. Kruskal–Wallis significance tests followed by Dunn’s post-hoc tests are conducted for all four stages, using a significance level of 0.05.


\section{Results} 
\label{sec: results}

\subsection{Fixed time budget}
\label{subsec: takeover performance}

\begin{figure*}[htbp]
    \centering
    \begin{subfigure}[b]{0.66\columnwidth}
        \centering
        \includegraphics[width=\linewidth]{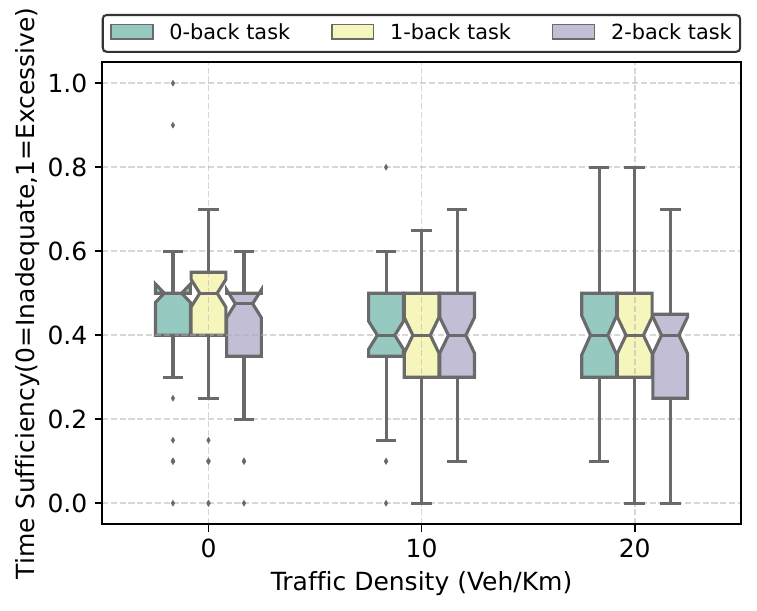}
        \caption{Perceived Time Sufficiency}
        \label{fig:subfig3}
    \end{subfigure}
    \hfill
    \begin{subfigure}[b]{0.66\columnwidth}
        \centering
        \includegraphics[width=\linewidth]{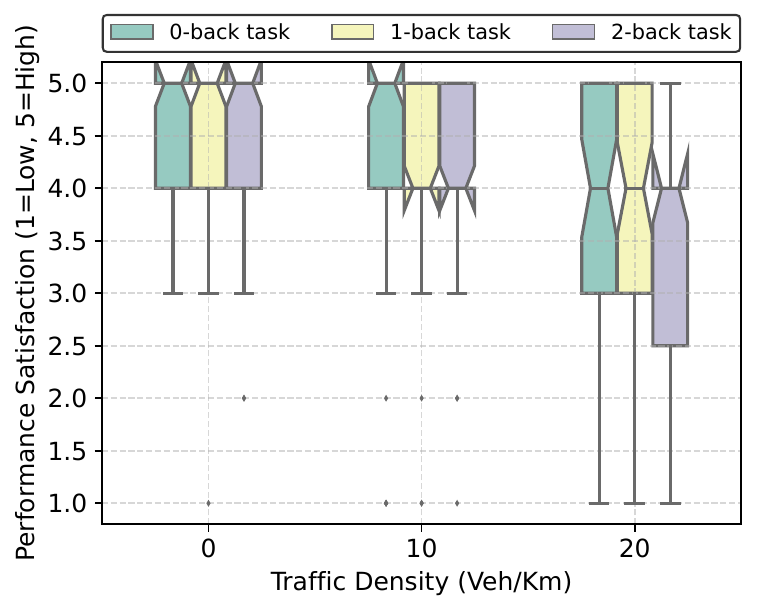}
        \caption{Perceived Performance Satisfaction}
        \label{fig:subfig3}
    \end{subfigure}
    \hfill
    \begin{subfigure}[b]{0.66\columnwidth}
        \centering
        \includegraphics[width=\linewidth]{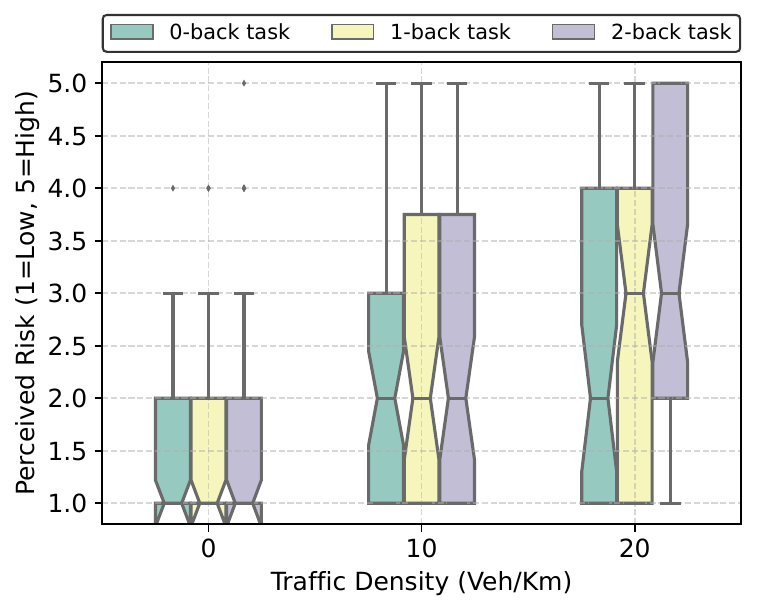}
        \caption{Perceived Risk}
        \label{fig:subfig2}
    \end{subfigure}

    \begin{subfigure}[b]{0.66\columnwidth}
        \centering
        \includegraphics[width=\linewidth]{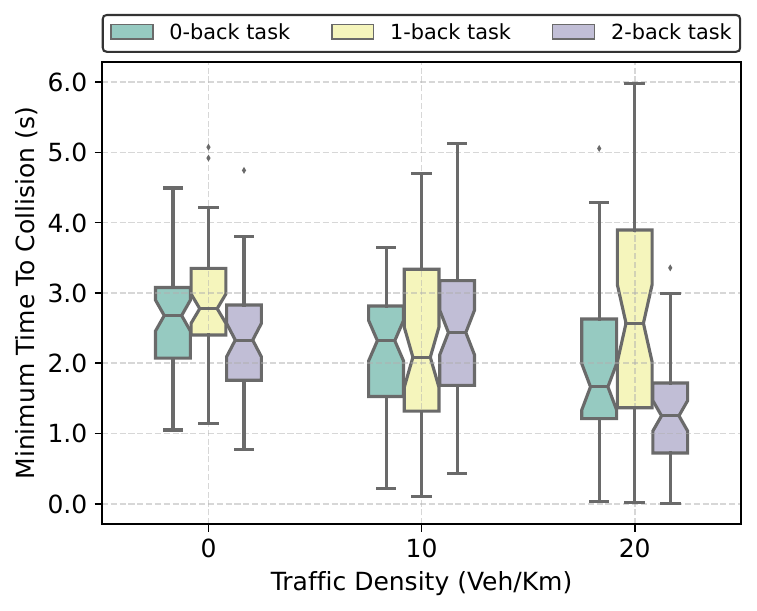}
        \caption{Minimum Time To Collision}
        \label{fig:subfig1}
    \end{subfigure}
    \hfill
    \begin{subfigure}[b]{0.66\columnwidth}
        \centering
        \includegraphics[width=\linewidth]{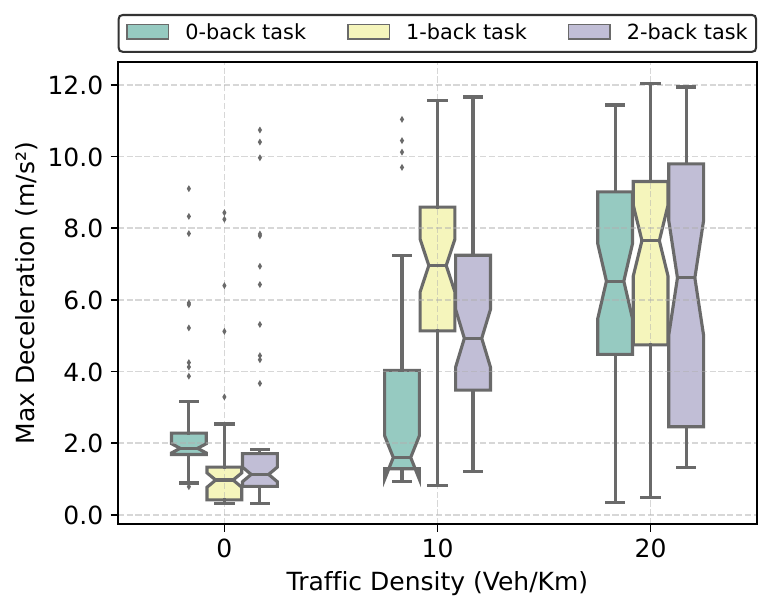}
        \caption{Max Deceleration}
        \label{fig:subfig3}
    \end{subfigure}
    \hfill
    \begin{subfigure}[b]{0.66\columnwidth}
        \centering
        \includegraphics[width=\linewidth]{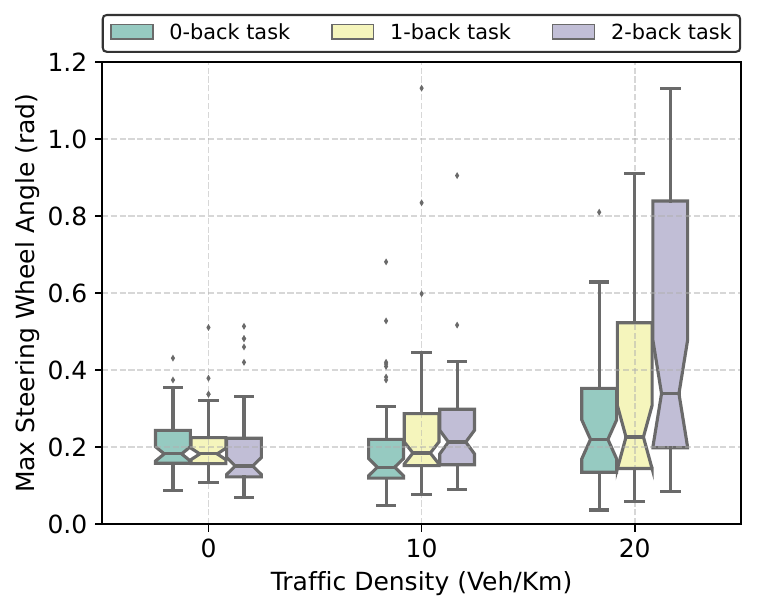}
        \caption{Max Steering Wheel Angle}
        \label{fig:subfig4}
    \end{subfigure}
    
    \caption{Takeover performance metrics across traffic densities and non-driving related tasks.}
    \label{fig: performance across scenarios}
\end{figure*}

To evaluate whether the fixed time budget offers sufficient and consistent support for safe and comfortable takeovers across diverse scenarios, this subsection examines drivers’ takeover performance across three traffic densities (0/10/20 vehicle/km) and three $n$-back tasks ($n = 0, 1, 2$) with a seven-second time budge. Specifically, three objective operational indicators are considered for measuring response quality: minimum Time-To-Collision (TTC), maximum deceleration, and maximum steering wheel angle. Three subjective driver perceptions are also included for capturing user experience: perceived time sufficiency, perceived performance satisfaction, and and perceived risk. Distributions of these metrics are illustrated in Figure \ref{fig: performance across scenarios}. Overall, with a fixed 7-second time budget, traffic density exerts a consistent and strong effect on both subjective experience and objective operations. Higher traffic density is generally associated with (i) worse driver experience, as indicated by reduced sense of time sufficiency ($H = 21.19, p < .001$), lower satisfaction ($H = 40.35, p < .001$), and increased perceived risk ($H = 50.46, p < .001$); and (ii) riskier evasive maneuvers, including shorter minimum TTC ($H = 31.27, p < .001$), greater maximum deceleration ($H = 155.98, p < .001$), and larger maximum steering wheel angles ($H = 24.29, p < .001$). In contrast, the $n$-back task shows more selective effects. It significantly lowers performance satisfaction ($H = 6.59, p = .037$), increases perceived risk ($H = 11.07, p = .004$), and reduces minimum TTC ($H = 18.30, p < .001$). However, it does not significantly alter the perceived time sufficiency ($p = .168$) and the intensity of control inputs, as reflected in maximum deceleration ($p = .147$) and steering angle ($p = .074$). These findings highlight that the effectiveness of a fixed time budget is highly context-dependent, and a uniform 7-second time budget often fails to meet drivers’ safety and comfort requirements, especially under complex conditions. They underscore the necessity of adaptive time budgets that adjust to both situational factors and individual needs to ensure driving safety and user experience.


\subsection{Buffer time allocation}
\label{subsec: takeover buffer}

\begin{figure*}[thpb]
    \centering
    \begin{subfigure}[b]{0.66\columnwidth}
        \centering
        \includegraphics[width=\linewidth]{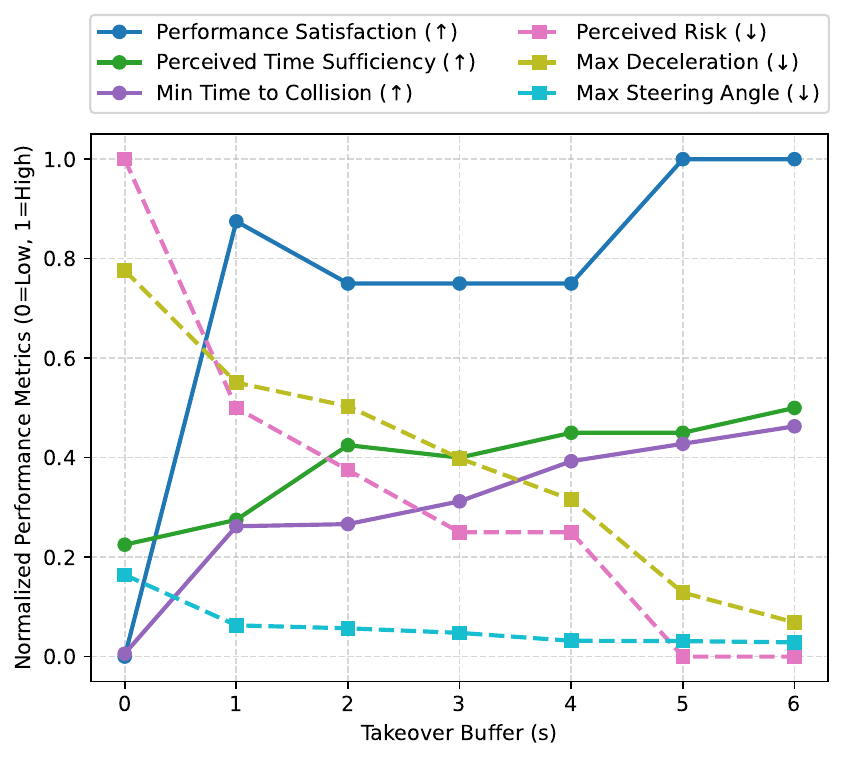}
        \caption{Takeover buffer}
        \label{subfig: performance_vs_tob}
    \end{subfigure}
    \hfill
    \begin{subfigure}[b]{0.66\columnwidth}
        \centering
        \includegraphics[width=\linewidth]{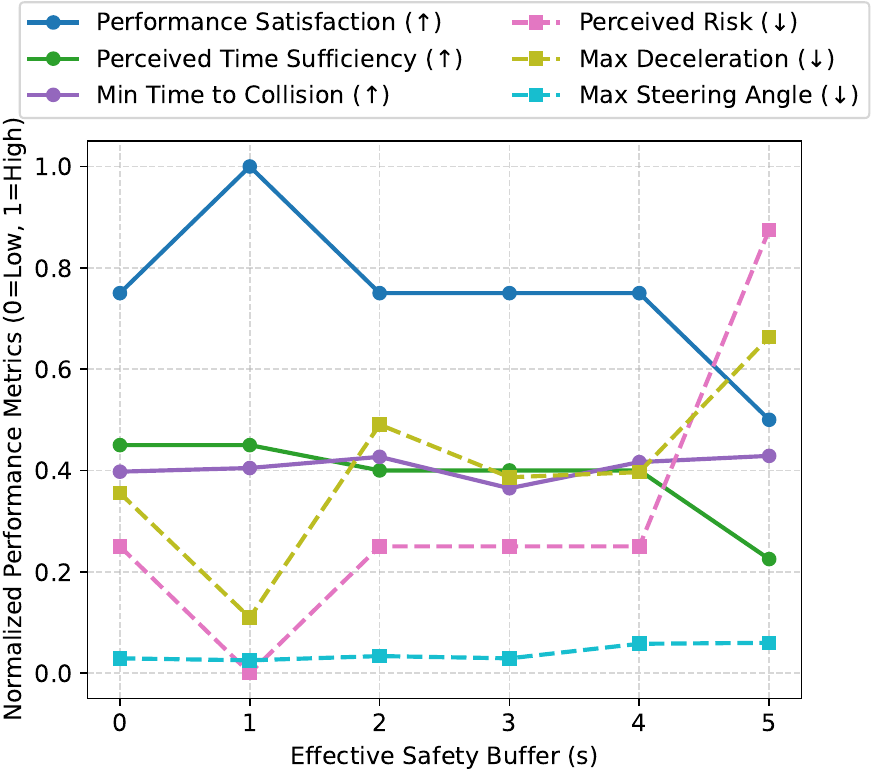}
        \caption{Effective safety buffer}
        \label{subfig: performance_vs_sb}
    \end{subfigure}
    \hfill
    \begin{subfigure}[b]{0.66\columnwidth}
        \centering
        \includegraphics[width=\linewidth]{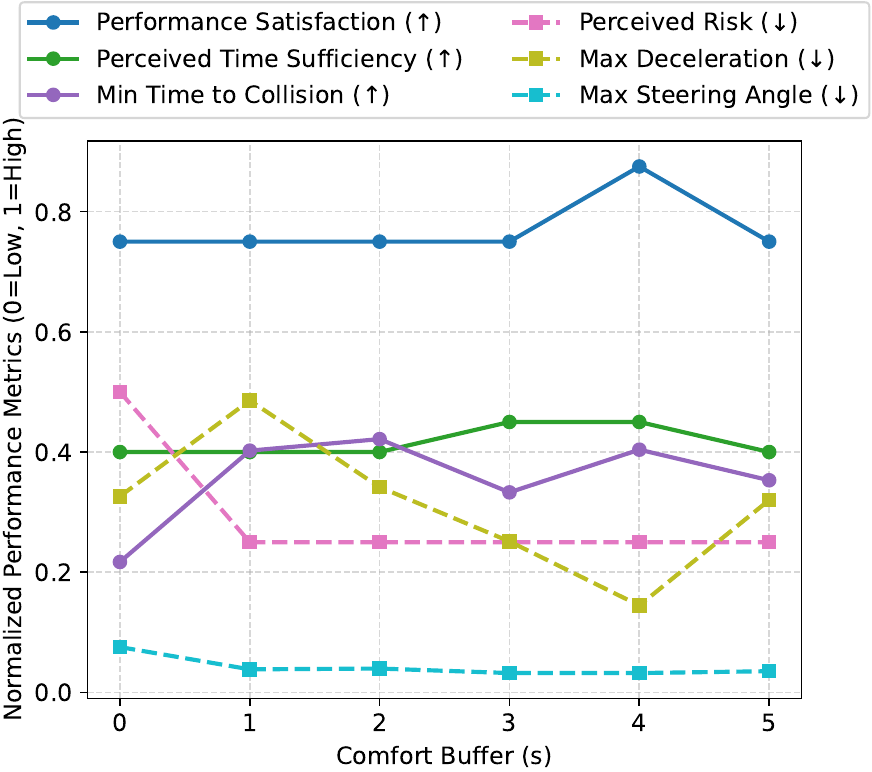}
        \caption{Comfort buffer}
        \label{subfig: performance_vs_cb}
    \end{subfigure}
    
    \caption{Effects of different buffer types on performance metrics ($\uparrow$: higher is better; $\downarrow$: lower is better).}
    \label{fig: performance_vs_buffers}
\end{figure*}

Time budget defines the total time available for drivers to resume conscious manual control once the CADS issues a takeover request. However, completing a takeover within this budget does not necessarily guarantee safety or comfort. A more informative measure is the takeover buffer, which represents the residual time margin remaining after the driver has successfully taken control. The takeover buffer can be decomposed into two complementary components: (i) the safety buffer, which represents the time margin required for driver intervention to ensure collision avoidance, and (ii) the comfort buffer, which provides additional time for drivers to regain control with a positive experience. In this sections, we analyze how variations in takeover buffers, together with their safety and comfort components, relate to both objective performance and subjective evaluations, as illustrated in Figure \ref{fig: performance_vs_buffers}. Solid lines indicate metrics where higher values are desirable; dashed lines indicate the opposite.

Generally, the takeover buffer significantly affects all six performance metrics ($p < .001$), as shown in Figure \ref{subfig: performance_vs_tob}. For the subjective measures: (i) performance satisfaction increases sharply from 0 to 1s, stabilizes around 0.75, and reaches a maximum plateau at 5–6s, with larger buffers yielding higher satisfaction than shorter ones ($H = 52.35, p < .001$); (ii) perceived time sufficiency shows a fluctuating upward trend as takeover buffer increases ($H = 26.32, p < .001$), but consistently remains below 0.5, indicating that drivers generally felt the takeover buffer was insufficient and desired more time; (iii) perceived risk decreases gradually with longer takeover buffer ($H = 46.37, p < .001$), reaching near 0 at 5–6s of takeover buffer. For the objective measures: (i) minimum TTC generally increases with longer takeover buffers ($H = 44.61, p < .001$), with no significant differences between the 4–6s conditions ($p > .05$); (ii) maximum deceleration decreases steadily as the takeover buffer increases, showing a consistent downward trend ($H = 82.65, p < .001$); (iii) maximum steering wheel angle is significantly higher for short takeover buffers than for longer ones ($H = 60.42, p < .001)$, with no significant differences observed between the 4–6 s conditions ($p > .05$). Overall, longer takeover buffers are associated with better performance, with 5s and 6s buffers producing nearly identical effects.




Within the takeover buffer, the safety buffer allows drivers to complete necessary evasive operations after regaining conscious control of the vehicle. Determining an appropriate duration for the safety buffer requires understanding how much time drivers actually need and how this duration affects various takeover performance metrics. In this study, all drivers successfully took over vehicle control without any accidents, satisfying the minimum safety requirements. The interval from when drivers consciously resume control of the vehicle until no further evasive maneuvers are performed is defined as the \textit{Effective Safety Buffer}. Intuitively, more challenging takeover scenarios would require drivers to take longer for evasive maneuvers, resulting in larger effective safety buffers. However, Kruskal–Wallis tests indicate that neither traffic density nor engagement in non-driving-related tasks significantly affect the effective safety buffer required by drivers ($p > .05$). Similarly, no significant association is observed with drivers’ ToT ($p = .07$). This underlines the need to consider the safety buffer as a robust design parameter, independent of situational variations. The effects of effective safety buffers on six performance metrics are presented in Figure \ref{subfig: performance_vs_sb}. Generally, a longer required safety time for evasive maneuvers is associated with (i) worse subjective experience, reflected by lower performance satisfaction ($H = 24.74, p < .001$), less perceived time sufficiency ($H = 14.49, p = .025$), and higher perceived risk ($H = 32.03, p < .001$); and (ii) more abrupt takeover maneuvers, as indicated by larger maximum deceleration ($H = 18.55, p = .005)$), lager maximum deceleration ($H = 54.04, p < .001$), and greater maximum steering angle ($H = 39.49, p < .001$). Notably, when the effective safety buffer that drivers require is shorter than 1s, drivers’ performance is suboptimal, likely due to rushed control or overreactions. This suggests that, in our experimental setting, a safety buffer longer than 1s is necessary to ensure safe and stable takeovers.





 Meanwhile, the allocated time budget should not only be sufficient to avoid accidents but also to enable a comfortable takeover experience. Therefore, we introduce the concept of the comfort buffer, which represents an additional, emotionally oriented margin of time that does not require takeover-related cognitive or operational activities. The effects of the comfort buffer on six performance metrics are shown in Figure \ref{subfig: performance_vs_cb}. A significant effect is found for maximum deceleration ($H = 22.893, p < 0.001, $), with a 4-second comfort buffer associated with less abrupt braking during takeovers. No significant effects are observed for the other five performance metrics ($p>0.05$).










\subsection{Driver-preferred time allocation}
\label{subsec: driver preference}

\begin{figure}[thpb]
    \centering
    \includegraphics[width=0.98\columnwidth]{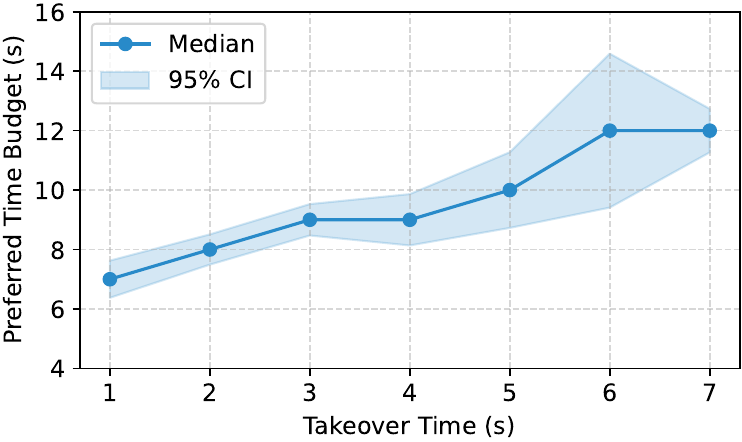}
    \caption{Drivers' preferred time budget by takeover time.}
    \label{fig: pre tb by tot}
\end{figure}

In addition to examining the effects of the provided time budget on takeover performance, we also consider drivers’ preferences for adjusting time budgets across different scenarios, which is essential for user acceptance and personalized automated driving. Figure \ref{fig: pre tb by tot} illustrates drivers' preferred time budgets across different ToT intervals. 

Overall, drivers generally prefer time budgets longer than the fixed 7s used in this study. Considering that all takeovers were completed safely, such preferences for longer time budgets therefore represent a desire for reduced stress and smoother transitions, rather than an indication of safety insufficiency. Meanwhile, drivers tend to request longer time budgets when more time is needed to return to the conscious driving loop ($H = 25.57, p < .001$). The relationship between median takeover time ($\overline{T\!oT}$) and drivers’ preferred time budgets ($pTB$) is well captured by a linear regression model ($R^2$=0.95, RMSE=0.40):

\begin{equation}
pTB = 0.86 * \overline{T\!oT} + 6.14
\end{equation}

This study further examines drivers' preferred takeover buffer across different ToT groups. Although the Kruskal–Wallis test indicates an overall significant difference across ToT groups ($H = 23.35, p<.05$), Dunn’s post-hoc comparisons reveals no significant differences within 1-2s groups (median: 6.35s, $p>0.05$), nor among the 3–7s groups (median: 5.30s, $p>0.05$). These results suggest that drivers’ preferred takeover buffer is largely stable across a wide range of ToTs.





\section{Discussion} 
\label{sec: discussion}

\subsection{Adaptive time budget}
\label{subsec: adaptive time budget}

This study demonstrates that the sufficiency of fixed time budgets is strongly context-dependent. Results show that a 7-second time budgets can be effective for simple scenarios or skilled drivers with takeover buffer exceeding 5s, but may impair user comfort and operation quality in complex scenarios or for less capable drivers when the takeover buffer falls below 5s. Similar findings have been reported in prior studies. For example, Feldhütter et al. \cite{feldhutter2019effect} reported that under a fixed 5s time budget, fatigued drivers exhibited lower takeover quality—characterized by higher decelerations, inappropriate trajectories and initial responses to the TOR, and increased crash risk—compared to alert drivers. Jin et al. \cite{jin2021modeling} found that in both 7s and 10s time budget conditions, higher traffic density led to more abrupt acceleration and deceleration inputs and less stable vehicle trajectories. These findings highlight the necessity of adaptive time budgets that dynamically adjust to both scenario and driver demands, as also suggested by \cite{li2021adaptive, tanshi2022determination, sekadakis2025identifying}.

To address this need, we propose an adaptive time budget framework composed of two components: the predicted takeover time and the driver’s preferred takeover buffer. 
Compared with the approach proposed by Tanshi and Söffker \cite{tanshi2022determination}, which defines the time budget as the sum of takeover time and maneuver response time, our framework incorporates an additional comfort buffer to further enhance user experience. Within this framework, the predicted takeover time reflects how long an individual driver is expected to regain control under a given situation, and can be estimated using prediction models from prior work (e.g., \cite{ayoub2022predicting, liang2025predicting}). The preferred takeover buffer represents an additional margin of time that should not only exceed a minimum safety threshold to ensure collision avoidance, but also avoid unnecessary extension, as excessively long buffers can reduce attentiveness to takeover requests and impair takeover efficiency \cite{tanshi2022determination}. By combining these two components, the framework allows the system to dynamically tailor the time budget to drivers' demands under specific situations. These results are relevant to safety and adaptive variation of the time budget for successful takeover.



Regarding the takeover buffer, our results suggest that a minimum threshold of at least one second is necessary to ensure stable and safe takeovers. This is because when the effective safety buffer dropped below one second, drivers often displayed rushed control and overreactions, which were reflected in reduced maneuver smoothness and lower subjective ratings. This lower bound aligns well with Eriksson et al. \cite{eriksson2017takeover}, who recommended a following gap of one car length per 16 km/h, translating to approximately 0.9 seconds at 100 km/h, and with Papadimitriou et al. \cite{papadimitriou2024method}, who identified a similar 0.9-second critical threshold. Importantly, however, our analysis also suggests that ``longer is not always better.'' Performance improvements plateau when the takeover buffer exceeds five seconds, with both 5-second and 6-second conditions yielding nearly identical optimal outcomes. This points to a saturation effect, in which additional buffer time provides limited added value. Such diminishing returns have also been observed in \cite{wan2018effects}. Moreover, overly long buffers may even impair driver performance by lowering the perceived urgency and reducing takeover efficiency. This was also pointed out by Li et. al \cite{li2024does} as they found that appropriate time pressure can reduce takeover time and improve takeover performance.

An interesting finding of this study is that drivers’ preferred takeover buffers remain relatively stable across different takeover time levels. Intuitively, more complex scenarios should increase the demand for larger buffers, yet our results show stable preferences: if drivers resume vehicle control within 3s, there is a large chance that they responded hastily or overreacted, as indicated by drivers' preference for slightly longer buffers (around 6.35s). Beyond this threshold, drivers appeared sufficiently re-engaged with an additional buffer of about 5.30s, showing little inclination for longer buffers to perform evasive maneuvers comfortably and safely. This stability may stem from the uniform evasive task across conditions (a single lane change), making buffer preferences more sensitive to operational complexity than to takeover duration. Another explanation is that buffer preferences are more strongly influenced by stable individual characteristics rather than situational demands. Our prior work \cite{liang2025multidimensional}, for instance, revealed systematic gender differences in perceived time sufficiency, with females reporting greater sufficiency than males under identical conditions. Similar individual-level factors may shape stable buffer preferences. These considerations underscore that, even though drivers’ preferred takeover buffer appears stable, the findings do not diminish the need for adaptive time budgets, since drivers may face evasive maneuvers of varying complexity and individual differences are likely to shape buffer requirements. Even if the preferred buffer itself is stable, combining variable takeover time predictions with this constant buffer can still yield an effective adaptive budget tailored to drivers and contexts.

Accordingly, we conduct a preliminary validation of our proposed adaptive strategy by comparing a fixed 7-second time budget with a piecewise takeover buffer function (6.35s when takeover time $\leq$ 3s, 5.30s otherwise). Results indicate that the adaptive strategy aligns significantly better with drivers’ preferred time budgets than the fixed 7-second allocation, with a 13\% reduction in MAE and a 16\% reduction in RMSE ($p < 0.01$). This suggests that even simple adaptive mechanisms can outperform fixed time budgets. In different takeover scenarios, the saturation value of the optimal buffer may vary depending on the specific demands of the maneuver and the driver. Future research should extend this validation by explicitly modeling operational complexity and driver characteristics, thus refining and optimizing adaptive time budget strategies across a wider range of contexts.




\subsection{Limitation}

This study has the following limitations. First, the experiment was conducted in a driving simulator, which, while providing a controlled and repeatable environment, may not fully capture the complexity and unpredictability of naturalistic driving. 
Real-world validation with a wider variety of evasive maneuvers and more diverse drivers is therefore essential to confirm the generalizability of the findings. Second, we argue that the proposed adaptive time budget framework primarily targets non-time-critical scenarios. Because in time-critical situations, such as sudden cut-ins, the CADS should trigger takeover requests or execute evasive actions immediately, leaving little room for designing a sufficient time budget. This study used a fixed 7-second time budget, which is generally considered as non-time-critical scenarios. However, drivers frequently preferred additional time to regain control safely and comfortably, indicating the need to validate the framework in scenarios with time budgets exceeding 7s. Third, learning effects may have influenced driver behaviors across repeated trials, potentially affecting the time they took to resume conscious vehicle control and their preferences for takeover buffers. To mitigate this, participants completed a 10-minute practice drive, scenarios were counterbalanced via a Latin Square design, and takeover requests were randomized between 30–60s in each session. Nonetheless, residual learning or anticipation effects may persist and should be considered in future studies. Fourth, although we conceptually validated our adaptive framework, participants did not directly experience dynamically adjusted time budgets. Future research should implement adaptive systems that adjust time budgets based on predicted takeover time and driver-preferred buffers. This would enable assessment of how adaptive time budgets affect takeover safety and subjective experience in realistic and interactive scenarios.





\section{Conclusion}
\label{sec: conclusion}


This study proposes an adaptive time budget framework as an important first step toward providing drivers with sufficient time to resume vehicle control across diverse scenarios. Specifically, we examine the influence of the takeover buffer on multiple performance metrics and incorporate drivers’ time budget preferences into the framework to support both user comfort and driving safety. Results indicate that takeover buffers of five to six seconds consistently yield high and stable levels of performance, suggesting that this range may represent an optimal window for balancing safety and comfort. This also indicates a potential saturation point, where further extending the buffer provides limited additional benefit and, in some cases, may even diminish subjective evaluations such as perceived satisfaction. These findings highlight the importance of considering not only sufficient buffer duration for safety but also the diminishing returns and possible drawbacks of excessively long buffers in time budget design. Within our experimental settings, drivers prefer relatively stable takeover buffers in addition to their required takeover times, regardless of varying traffic densities and $n$-back tasks. Therefore, the adaptive time budget framework integrates predicted takeover time (pToT, estimated using data-driven models) with a preferred takeover buffer, represented as a piecewise rule: when the pToT is shorter than three seconds, a buffer of 6.35s is added; when the pToT exceeds three seconds, a buffer of 5.30s is applied. Primary validation shows that the time budgets generated by this framework are better aligned with driver preferences compared with the fixed time budget while ensuring minimum safety requirements.

In summary, this study has three main contributions:
\begin{itemize}
    \item provides an empirical analysis of the relationship between takeover buffer, takeover time, and multiple dimensions of takeover performance (integrating both objective safety metrics and subjective driver evaluations);

    \item reveals a consistent preference for takeover buffer across different takeover time conditions, underscoring a stable driver requirement for supplementary time to facilitate safe and comfortable control transitions; and

    \item proposes an adaptive time budget allocation strategy—predicted takeover time plus a saturated takeover buffer—to accommodate situational demands in conditionally automated driving.
\end{itemize}

Our findings offer a foundation for designing flexible, human-centered takeover strategies in conditionally automated driving, promoting both safe and comfortable driver–automation interactions. Future research is recommended to (i) investigate how operation complexity and driver characteristics influence preferred takeover buffers, and (ii) experimentally test the adaptive time budget framework in diverse and naturalistic driving environments to confirm its effectiveness and generalizability. 



%





\ifCLASSOPTIONcaptionsoff
  \newpage
\fi



\bibliographystyle{IEEEtran}
\bibliography{IEEEexample}


%


\begin{IEEEbiography}[{\includegraphics[width=1in,height=1.25in,clip]{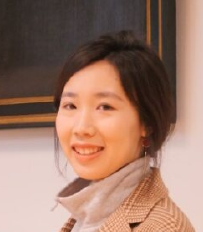}}]{Kexin Liang}
is a PhD candidate in the Department of Transport \& Planning at Delft University of Technology (TU Delft), Delft, the Netherlands. She received her master in control science and engineering from Beijing Jiaotong University in 2021.
\end{IEEEbiography}

\begin{IEEEbiography}[{\includegraphics[width=1in,height=1.25in,clip]{simeon.pdf}}]{Simeon C. Calvert}
is an associate professor of Smart \& Automated Driving in the department of Transport \& Planning at the TU Delft. He is the director and founder of the Automated Driving \& Simulation (ADaS) research lab and co-director of the CityAI-lab for research on urban behaviour using AI. He is also a board member of the Centre for Meaningful Human Control over AI. He received his PhD in probabilistic macroscopic traffic flow modeling from TU Delft in 2016.
\end{IEEEbiography}

\begin{IEEEbiography}[{\includegraphics[width=1in,height=1.25in, clip]{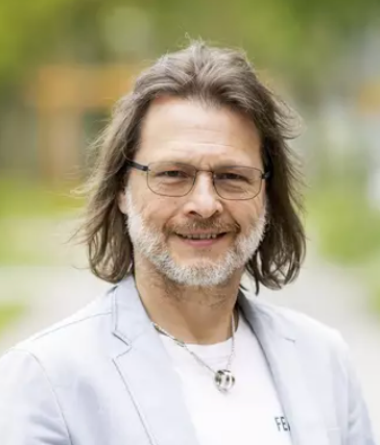}}]{J.W.C. van Lint}
is a full professor of Traffic Simulation and Computing in the department of Transport \& Planning at the TU Delft. He received the M.Sc. degree in civil engineering informatics and the Ph.D. degree in transportation from TU Delft in 1997 and 2004, respectively. After receiving the Ph.D. degree, he was appointed as an Anthonie van Leeuwenhoek (AvL) Full Professor in 2013. He is currently the Director of the Digitisation \& AI For Mobility Network Dynamics Laboratory (DAIMoND Lab) in the Faculty of Civil Engineering and Geosciences. 
\end{IEEEbiography}
%




\end{document}